\documentclass[11pt,onecolumn]{article}
\usepackage[utf8]{inputenc}

\usepackage{mathtools}
\usepackage{amsmath}
\usepackage{amssymb}
\usepackage{amsthm}
\usepackage{amsfonts}
\usepackage{geometry}
\usepackage{graphicx}
\usepackage[dvips]{epsfig}
\usepackage{braket}
\usepackage[affil-it]{authblk} 
\usepackage{hyperref}
\usepackage{cite}
\usepackage{placeins}

\begin{document}
\title{Quantization of Length in Spaces with Position-Dependent Noncommutativity}
\author{Jishnu Aryampilly\thanks{\href{mailto:jishnu.sankaran@pondiuni.ac.in}{jishnu.sankaran@pondiuni.ac.in}}}
\author{Muthukumar Balasundaram\thanks{\href{mailto:muthukbs@pondiuni.ac.in}{muthukbs@pondiuni.ac.in}}}
\author{Aamir Rashid\thanks{\href{mailto:aamirjamian@pondiuni.ac.in}{aamirjamian@pondiuni.ac.in}}}

 \affil{Department of Physics, \\Pondicherry University, \\ Puducherry 605014, India}

\date{Dated: \today}
\maketitle

\begin{abstract}

We present a novel approach to quantizing the length in noncommutative spaces with positional-dependent noncommutativity. The method involves constructing ladder operators that change the length not only along a plane but also along the third direction due to a noncommutative parameter that is a combination of canonical/Weyl-Moyal type and Lie algebraic type. The primary quantization of length in canonical-type noncommutative space takes place only on a plane, while in the present case, it happens in all three directions. We establish an operator algebra that allows for the raising or lowering of eigenvalues of the operator corresponding to the square of the length. We also attempt to determine how the obtained ladder operators act on different states and work out the eigenvalues of the square of the length operator in terms of eigenvalues corresponding to the ladder operators. We conclude by discussing the results obtained.

\end{abstract}

\section{Introduction}
In the expansive realm of fundamental physics, the notion of spacetime stands as a bedrock upon which our understanding of the universe is constructed. Conventionally, the principles of classical physics have offered a framework to describe spacetime where space and time coordinates exhibit commutativity, enabling precise measurements of position and duration. Nevertheless, emerging theories and frameworks have heralded a departure from this classical paradigm, revealing the intriguing prospects of a noncommutative spacetime. The idea of noncommutativity in the coordinates of spacetime can be attributed to Heisenberg \cite{Jackiw:2001dj}, and it was subsequently developed further by Snyder \cite{Snyder:1946qz} as a means to address the issue of divergences in quantum field theory. Noncommutative spacetime theory suggests that space and time coordinates do not commute but instead exhibit a fundamental uncertainty or noncommutativity relation. This revelation necessitates acknowledging that precise measurements of both position and time are inherently uncertain at infinitesimal scales. This departure from classical notions has sparked considerable interest and led to the formulation of various theoretical models attempting to capture the essence of noncommutative spacetime \cite{Banerjee:2009gr, Padmanabhan:2012wha, Madore:1999bi}. The growing interest in noncommutative spaces is connected to the prediction of noncommutative structures in string theory and loop quantum gravity \cite{Seiberg:1999vs, Addazi:2017bbg, Frohlich:1993es, Frohlich:1995mr}. The literature on noncommutative theories is replete in the contexts of quantum field theories \cite{Szabo:2001kg, Gopakumar:2000zd, Joseph:2008fz, Mandal:2001xv, Girotti:2003at, Douglas:2001ba, Grensing:2013leq, Balachandran:2010wc, Doplicher:1994tu, Alvarez-Gaume:2006qrw, Filk:1996dm}, quantum mechanics \cite{Saha:2010zh, Nair:2000ii, Muthukumar:2006ab, Muthukumar:2007zza, Sinha:2011dt, Biswas:2019obt, Bolonek:2002cc, Aschieri:2005yw, Muhuri:2020did, Harikumar:2004qc} and gravity theories \cite{Calmet:2005qm, Banerjee:2007th, Balachandran:2006qg, Harikumar:2006xf, Roy:2022rav, Chaichian:2007we}.

In the case of noncommutative geometry by Alain Connes \cite{Connes:1997cr, Connes:2000ti}, the spectral manifold is shown to have a geometric analog of the Heisenberg commutation relation involving the Dirac operator and Feynman slash of real scalar fields, leading to the quantization of volume \cite{Chamseddine:2014nxa}. The idea of length as an operator has already been discussed in canonical quantum gravity \cite{Thiemann:1996at}. 
Loop quantum gravity has developed the rigorous construction of (spatial) geometrical operators, such as the area and the volume\cite{Rovelli:1994ge, Ashtekar:1997fb}.

Within the context of noncommutative spaces, the idea of length as an operator was proposed in \cite{Balasundaram:2022esk}, where it was shown to result in the quantization of length in the  noncommutative space with the canonical/Weyl-Moyal type noncommutativity 
\begin{equation}
[\hat{x}^{\mu},\hat{x}^{\nu}]=i\theta^{\mu\nu}
\end{equation}
among the coordinates where $\theta^{\mu\nu}$ is a constant and real antisymmetric matrix. The motivation behind such a proposal was the following. The noncommutativity of spatial coordinates is closely linked to the existence of a minimum length within a system. This minimum length is commonly associated with inherent uncertainties in distance measurements. Rather than associating the minimum length with uncertainties, it can be attributed to the minimum value of the quantized length. If length is taken as an operator with a spectrum of eigenvalues, then a set of ladder operators may exist to go from one eigenstate to another such that
\begin{equation}
[\hat{L}^2,\hat{a}]=\lambda \,\hat{a}, \label{len-lad-op-comm0}
\end{equation}
where $\hat{L}^2$ is the operator corresponding to the square of length and $\hat{a}$ and $\hat{a}^{\dagger}$ being the ladder operators. If $ \hat{a}=\sum_\mu\alpha_\mu\,\hat{x}^\mu$, where $\alpha_\mu$'s are complex constants, then Eq.(\ref{len-lad-op-comm0}) leads to an eigenvalue equation which, in the case of 3-D with $\theta^{12}=\theta^{13}=\theta^{23}=\theta$, gives three real eigenvalues and their corresponding eigenvectors. Two of the eigenvectors give the required ladder operators that change the length in the plane formed by them. The third eigenvector points in the direction normal to that plane and the length is not quantized along this normal direction\cite{Balasundaram:2022esk}. 

In this paper, we follow an approach similar to the operator methods in the quantum harmonic oscillator and the angular momentum problems and apply it to the case of position-dependent noncommutativity. Position-dependent noncommutativity has been discussed before in \cite{Lawson:2020vbd, Blaschke:2018vsl, Gayral:2005ih, Gomes:2009tk, Fring:2010pw}. In particular, the noncommutative parameter used in our approach is the following combination of the canonical/Weyl-Moyal type and Lie algebraic type:
\begin{equation}
[\hat{x}^\mu,\hat{x}^\nu]=i(\theta^{\mu\nu}+B^{\mu\nu}_{\phantom{~~}\rho} \, \hat{x}^{\rho}), 
\label{commutator1}
\end{equation}
where $\theta^{\mu\nu}$ corresponds to a constant and real antisymmetric matrix and $ B^{\mu\nu\rho}$ is real and completely antisymmetric. 

The paper is organized as follows. In Section 2, we establish an operator algebra of the length-square operator that allows for the raising and lowering of eigenvalues of $\hat{L}^2$. In Section 3, we apply this approach in a 3-dimensional space. We construct a commutation relation between the operator $\hat{L}^2$ corresponding to the square of length and the ladder operators analogous to the commutation relation between the Hamiltonian of the harmonic oscillator and its raising/lowering operator.
We work with this commutation relation to obtain an eigenvalue equation and consequently construct a set of operators $\hat{a}_{-}$, $\hat{a}_{+}$ and $\hat{b}$. Once we have obtained the operators, we adopt them to construct a ladder of states that constitute the eigenstates of the $\hat{L}^2$ operator. In Section 4,  we work out the eigenvalues of $\hat{L}^2$ in terms of eigenvalues corresponding to the operators $\hat{a}_{+} \hat{a}_{-}$ and $\hat{b}$. The actions of $\hat{a}_{-}$ and $\hat{a}_{+}$ change not only the eigenvalues of $\hat{L}^2$ but also the eigenvalues of $\hat{b}$,  and in this way, the length is quantized along the direction of $\hat{b}$ also. In Section 5, by introducing another operator $\hat{K}$ that commutes with the ladder operators, we investigate the system's degeneracy further. We conclude in Section 6.

\section{Construction of Ladder Operators}

We establish an operator algebra in a manner that allows for the raising or lowering of eigenvalues of the operator $\hat{L}^2$. We define the operator corresponding to the square of the distance as 
\begin{equation}
\hat{L}^2=g_{\mu\nu}\hat{x}^{\mu}\hat{x}^{\nu}, \label{length-square1}
\end{equation}
where $g_{\mu\nu}$ is a constant symmetric metric of $D$-dimensional spacetime and Einstein's summation convention is used over the repeated indices $\mu$ and $\nu$ which take the values $1,2,\ldots D$. The prescription to construct a set of ladder operators $\{\hat{a}^{\mu}\}$ is that they satisfy the following commutation relation: 
\begin{equation}
[\hat{L}^2,\hat{a}^{\mu}]=\lambda  \, \hat{a}^{\mu},  \label{length-lad-comm}
\end{equation}
where $\lambda$ is a constant to be determined and $\hat{a}^{\mu}$ is linearly related to $\hat{x}^{\mu}$ as
\begin{equation}
\hat{a}^{\mu}=U^{\mu}_{\phantom{\mu}\nu}\,\hat{x}^{\nu}, \label{lad-op-exp1}
\end{equation}
where $U^{\mu}_{\phantom{\mu}\nu}$ is the transformation matrix. 
Substituting Eq.(\ref{length-square1}) and Eq.(\ref{lad-op-exp1}) into Eq.(\ref{length-lad-comm}) and using Eq.(\ref{commutator1}) leads to the following operator equation:
\begin{equation}
2i \,\theta_{\mu}^{\phantom{\mu}\rho}\,  U^{\sigma}_{\phantom{\mu}\rho}\,\hat{x}^\mu\,+\,i\,U^{\sigma}_{\phantom{\mu}\rho}\,B_{\nu\phantom{\rho}\kappa}^{\phantom{\nu}\rho}\,(\hat{x}^\nu\,\hat{x}^\kappa\,+\,\hat{x}^\kappa\,\hat{x}^\nu) =\lambda \, U^{\sigma}_{\phantom{\mu}\mu}\,\hat{x}^\mu. \label{U-eigen-op-eqn} 
\end{equation}
Since $(\hat{x}^\nu\,\hat{x}^\kappa\,+\,\hat{x}^\kappa\,\hat{x}^\nu)$ is symmetric  and $B_{\nu\phantom{\rho}\kappa}^{\phantom{\nu}\rho}$ is antisymmetric under the exchange of $\nu$ and $\kappa$, the second term is zero which leads to the eigenvalue equation for the transformation matrix: 
\begin{equation}
2i\,\theta_{\mu}^{\phantom{\mu}\rho}\,  U^{\sigma}_{\phantom{\mu}\rho}\, =\lambda \, U^{\sigma}_{\phantom{\mu}\mu}\,. \label{U-eigeneqn} 
\end{equation}

If
$
 \hat{X}^\dagger=
\begin{pmatrix}
\hat{x}^1, \hat{x}^2,\ldots \hat{x}^D
\end{pmatrix}
$ and $g$ is the matrix form of the metric tensor, then 
\begin{equation}
\hat{L}^2=\hat{X}^\dagger\,g\,\hat{X}.\label{len-sq-X}
\end{equation}
To relate $\hat{L}^2$ to the ladder operators we define $
 \hat{A}^\dagger=
\begin{pmatrix}
\hat{a}^{1\dagger}, \hat{a}^{2\dagger}, \ldots \hat{a}^{D\dagger}\\
\end{pmatrix}
$. Going by the analogy with harmonic oscillator and angular momentum problems, the ladder operators will be useful only if the length operator is related to number operators $(\hat{a}^{1}){}^\dagger\hat{a}^1, \,(\hat{a}^{2}){}^\dagger\hat{a}^2,\ldots$. Therefore, we require $\hat{L}^2$ in the following form:
\begin{equation}
\hat{L}^2=\frac{1}{\gamma}\, \hat{A}^\dagger g \hat{A}=\frac{1}{\gamma} \,\hat{X}^\dagger U^\dagger \,g\,U \hat{X}, \label{len-sq-AdaggerA}
\end{equation}
where  $\gamma$ is a constant. Comparing Eq.(\ref{len-sq-X}) and Eq.(\ref{len-sq-AdaggerA}), we get following condition for the transformation matrix:
\begin{equation}
U^\dagger \,g\,U = \gamma \, g.
\end{equation}

\section{The Length Operator in 3-D Space }

 We attempt to apply our approach to a 3-dimensional space. 
 For this case, the commutation relation that we use can therefore be defined as
 \begin{equation}
[\hat{x}^i,\hat{x}^j]=i(\theta^{ij}+B^{ij}_{\phantom{ij}k} \, \hat{x}_{k}), 
\label{commutator2}
 \end{equation}
 We define the ladder operator as
\begin{equation}
\hat{a}=\alpha_k\hat{x}^k,
\label{ladder_op}
\end{equation}
where Einstein's summation convention is implied and $\alpha_k$'s are complex constants to be determined. The operator corresponding to the square of the length, in this case, thus becomes
\begin{equation}
    \hat{L}^2=g_{ij}\hat{x}^{i}\hat{x}^{j}
    \label{length-square2}
\end{equation}
As discussed, we assume the following relation in analogy with the angular momentum operator in quantum mechanics
\begin{equation}
[\hat{L}^2,\hat{a}]=\lambda \hat{a} .
\label{len_sq_relation}
\end{equation}
Then, the substitution of Eq.(\ref{length-square2}) and Eq.(\ref{ladder_op}) in Eq.(\ref{len_sq_relation}) and using the commutator in Eq.(\ref{commutator2}) gives the relation
\begin{equation}
2i\theta^{k}_{\phantom{k}m}\alpha_{k}=\lambda\alpha_{m},
\label{eigen}
\end{equation}
where $\theta^{km}$ is a constant antisymmetric matrix and $g_{ij}$ is assumed to be diag(1,1,1).

We assume  the entries of $ \theta^{km} $ to be $ \theta^{12}=\theta^{13}=\theta^{23}=\theta$, which leads Eq.(\ref{eigen}) to give the nontrivial eigenvalues for $\lambda= \pm{2\sqrt{3}\theta}$. The third trivial solution is $\lambda=0$. 
The set of values for $\alpha_i$  corresponding to  $\lambda= -{2\sqrt{3}\theta}$ is worked out to be $(\alpha_1,\,\alpha_2,\,\alpha_3)= 
 (\rho\sigma, -\rho\sigma^*\rho)$, where $\rho= \displaystyle e^{i\delta_1}$ and $\sigma=-e^{i\pi/3}$. The operator $\hat{a}$ corresponding to this negative $\lambda$ is denoted by $\hat{a}_{-}$. It is identified with the lowering operation by comparing Eq.(\ref{len_sq_relation}) with a negative $\lambda$ to an analogous relation in the harmonic oscillator problem. The lowering operator can then be expressed as 
\begin{eqnarray}
 \hat{a}_-=\rho \left[\sigma \hat{x}^1- \sigma^* \hat{x}^2+\hat{x}^3\right].  
\end{eqnarray}
The eigenvector for the positive value $\lambda={2\sqrt{3}\theta}$ leads to the raising operator $\hat{a}_+=(\hat{a}_-)^\dagger$. The eigenvector corresponding to the third eigenvalue, that is, the trivial solution $\lambda=0$, leads to the operator $\hat{b}=\beta_i\,\hat{x}^i$ with
$(\beta_1,\beta_2,\beta_3)$ = $(1,-1,1)$. With these denotations, we write the Hermitian conjugate of the basis $\hat{A}$ in Eq.(\ref{len-sq-AdaggerA}) as 
$\hat{A}^\dagger=(\hat{a}_+, \hat{a}_-, \hat{b})$. 
The explicit values of $ \alpha_{i}$ in Eq.(\ref{eigen}) and $\beta_{i}$ have the properties 
such as $\alpha_i\alpha^i=0$, $\alpha_i\,\beta^i=\alpha_i^{*}\,\beta^i=0$ and $\theta^{ij}\beta_j=0$. 

Essentially, the three eigenvalues for $\lambda$ in Eq.(\ref{len_sq_relation}) lead to following commutators
\begin{equation}
    [\hat{L}^2,\hat{a}_\pm]= \pm 2\sqrt{3}\theta\hat{a}_\pm, 
    \label{len-sq-apm}
\end{equation}
and
\begin{equation}
    [\hat{L}^2,\hat{b}]= 0. 
    \label{len-sq-b}
\end{equation}
Also, the commutation relations among the ladder operators $\hat{a}_-$, $\hat{a}_+$ and $\hat{b}$ are obtained as 
\begin{align}
   [\hat{a}_-,\hat{a}_+]&=\sqrt{3}(3\theta+B\hat{b}),
   \label{aa-commutator}\\
  [\hat{b},\hat{a}_\pm]&=\mp\sqrt3\,B\,\hat{a}_\pm , 
  \label{ab-commutator}\\
 [\hat{a}_+\hat{a}_-,\hat{a}_+]&=\sqrt{3}(3\theta+B\hat{b}+\sqrt{3}B)\hat{a}_+,
   \label{aaa+commutator}\\
[\hat{a}_+\hat{a}_-,\hat{a}_-]&= -\sqrt{3}(3\theta+B\hat{b})\hat{a}_-,
   \label{aaa-commutator}
  \end{align}
With these commutators, the operator form of the square of length is expressed as,
\begin{equation}
\hat{L}^2=\frac{1}{3}[2\hat{a}_{+}\hat{a}_{-}+\sqrt3(3\theta+B\hat{b})+\hat{b}^{2}].
\label{lensq-op1}
\end{equation}
The Eq.(\ref{len-sq-apm}) leads to $[\hat{L}^2, \hat{a}_+\hat{a}_-]=0$ and since $[\hat{L}^2, \hat{b}]=0$, it is possible to construct a complete set of simultaneous eigenstates of $\hat{L}^2$, $\hat{a}_+\hat{a}_-$ and $\hat{b}$. 

\section{Eigenvalues of the Length-Square Operator}

Let us start with an eigenstate $|n\rangle$ of the number operator $\hat{a}_+\hat{a}_-$ and consider the action of the ladder operators on it. From Eq.(\ref{aaa+commutator}) and Eq.(\ref{aaa-commutator}), it is clear that the action of $\hat{a}_\pm$ on $|n\rangle$ is to raise/lower the eigenvalue of $\hat{a}_+\hat{a}_-$. So, we define the actions of $\hat{a}_{-}$, $\hat{a}_{+}$ and $\hat{b}$ respectively on the normalized state $|n\rangle$ as
\begin{align}
    \hat{a}_{-}|n\rangle &= h_{1}(n)|n-1\rangle,
    \label{h1}\\
    \hat{a}_{+}|n\rangle &= h_{2}(n)|n+1\rangle,
    \label{h2}\\
     \hat{b}|n\rangle &= g(n)|n\rangle.
     \label{g(n)}
\end{align}
Considering the Hermitian conjugate of Eq.(\ref{h1}), we obtain 
\begin{equation}
   \langle n|\hat{a}_{+} = \langle n-1|h_{1}^{*}(n).
   \label{h1_conj}
\end{equation}
Thus, it can easily be shown that
\begin{equation}
    \langle n|\hat{a}_{+}\hat{a}_{-}|n\rangle = \langle n-1|h_{1}^{*}(n) h_{1}(n)|n-1\rangle = {|{h_{1}(n)}|}^2.
    \label{h1-sq}
\end{equation}
Similarly upon taking the Hermitian conjugate of Eq.(\ref{h2}), we can obtain
\begin{equation}
     \langle n|\hat{a}_{-}\hat{a}_{+}|n\rangle = {|{h_{2}(n)}|}^2.
     \label{h2-sq}
\end{equation}
It can be further rewritten in terms of commutator to obtain the relation  $\langle n| \hat{a}_-\hat{a}_+|n\rangle = \langle n|( \hat{a}_+\hat{a}_- + [\hat{a}_-, \hat{a}_+])|n\rangle$ and using the commutator in Eq.(\ref{aa-commutator}), we obtain the relation
\begin{equation}
\langle n|\hat{a}_-\hat{a}_+|n\rangle
= \langle n|(\hat{a}_+\hat{a}_- +3\sqrt{3}\theta + \sqrt{3}B\hat{b})|n\rangle.
\end{equation}
The above relation can be reexpressed using Eq.(\ref{h1-sq}), Eq.(\ref{h2-sq}) and Eq.(\ref{g(n)}) as
\begin{equation}
|h_{2}(n)|^2=|h_{1}(n)|^2+3\sqrt3\theta+\sqrt{3}Bg(n)
\label{h2-h1}
\end{equation}
Considering the Hermitian conjugation of Eq.(\ref{h1}) with $n+1$ in place of $n$, we have $\langle n+1|\hat{a}_+=\langle n|{h_1}^*(n+1)$. Therefore, $\langle n |\hat{a}_+|n\rangle$ gives ${h_1}^*(n+1)$ on one hand but on the other hand it gives ${h_2}(n)$, which leads to the relation
\begin{equation}
|h_{2}(n)|^2=|h_{1}(n+1)|^2 .
\label{h2-h1(n+1)}
\end{equation}
Also, considering the action of $\hat{b}$ on the state $\hat{a}|n\rangle$ and by using the commutator in Eq.(\ref{ab-commutator}), we can find the relation between $g(n+p)$ and $g(n)$ for any integer $p$
\begin{equation}
    g(n+p)=g(n)-pB\sqrt3 .
    \label{g(n+m)}
\end{equation}
The Eq.(\ref{len-sq-apm}) implies that the eigenvalues of $\hat{L}^2$ are decreased by $\hat{a}_-$ by the amount $2\sqrt{3}\theta$. This decrease cannot go on forever since $\hat{L}^2$ will take negative values, which is unphysical. So, let us define a ground state of the system by $|\overline{n}\rangle$ such that the action of the lowering operator $\hat{a}_-$ on it gives 
\begin{equation}
\hat{a}_-|\overline{n}\rangle = 0.
\end{equation}
Here, we have $h_{1}(\overline{n})=0$ or
\begin{equation}
    |h_{1}(\overline{n})|^2=0 .
\end{equation}
We can further extend our analysis using Eq.(\ref{h2-h1}), Eq.(\ref{h2-h1(n+1)}) and Eq.(\ref{g(n+m)}) to compute values of $h_1$ as
\begin{align}
    |h_{1}(\overline{n}+1)|^2&=[3\sqrt3\theta+\sqrt3Bg(\overline{n})],\\
    |h_{1}(\overline{n}+2)|^2&= 2 [3\sqrt3\theta+\sqrt3Bg(\overline{n})]-3B^{2}  
\end{align}
and so on. The analysis can be further extended to obtain
\begin{equation}
|h_{1}(\overline{n}+m)|^2 = m[3\sqrt3\theta+\sqrt3B(g(\overline{n})-\frac{(m-1)}{2}\sqrt3B)] .
\label{h1(n+m)}
\end{equation}
For a general $n$, where $ n= \overline{n}+m$ and $m\geq 0$, the above equation modifies to a function involving $n$ and $\overline{n}$. 
In a similar manner, the general form of (\ref{g(n+m)}) is, then, better expressed as a function of $n$ and $\overline{n}$
\begin{equation}
    g(n)= g(\overline{n})-(n-\overline{n})\sqrt3B .
    \label{g(n)-2}
\end{equation}
where $g(\overline{n})$ comes from the operator $\hat{b}$ acting on the ground state, that is,
\begin{equation}
    \hat{b}|\overline{n}\rangle = g(\overline{n})|\overline{n}\rangle .
\end{equation}
Accordingly, Eq.(\ref{h1(n+m)}) can be rewritten as
\begin{equation}
    |h_{1}(n)|^2\!=\! (n-\overline{n})[3\sqrt3\theta+\sqrt3B(g(\overline{n})\!-\!\tfrac{(n-\overline{n}-1)}{2}\sqrt3B)].
    \label{h1(n)}
    \end{equation}
We can also find using Eq.(\ref{h2-h1(n+1)}) that
\begin{equation}
  {|h_2(n)|}^2=(n-\overline{n}+1)[3\sqrt3\theta+\sqrt3B(g(\overline{n})\!-\!\tfrac{(n-\overline{n})}{2}\sqrt3B)]
  \label{h2(n)}
\end{equation}
Since $|h_1(n)|^2$ and $|h_2(n)|^2$ cannot take negative values for any $n$, we can infer that
\begin{equation}
    3\sqrt{3}\theta + \sqrt{3}Bg(\overline{n}) \ge \frac{(n-\overline{n})}{2}3B^2.
    \label{g-in}
\end{equation}
Suppose $\tilde {n}$ is the maximum value $n$ can take such that the above inequality stands valid. Then, we can fix a top-most state  $|\tilde{n}\rangle$ such that
\begin{equation}
    \hat{a}_+|\tilde{n}\rangle = 0 .
\end{equation}
Now, since $h_2(\tilde{n}) = 0$ and $\tilde{n}\geq \overline{n}$, using (\ref{h2(n)}), we can express $g(\overline{n})$ in terms of $\tilde{n}$ and $\overline{n}$. Thus, we obtain
\begin{equation}
    g(\overline{n}) = \frac{(\tilde{n}-\overline{n})}{2}\sqrt{3}B - \frac{3\theta}{B}.
    \label{g(n-bar)}
\end{equation}
Putting Eq.(\ref{g(n-bar)}) back into Eq.(\ref{g-in}), we get $\tilde{n}\geq \overline{n}+m$. So essentially $n=\overline{n},~\overline{n}+1,~\ldots, \tilde{n}$ and $\tilde{n}=\overline{n},~\overline{n}+1,\ldots.$ But there is no restriction on $\overline{n}$, and it can take both positive and negative integer values.

We may now proceed to solve for the eigenvalues of $L^2$. The length-square operator in Eq.(\ref{lensq-op1}) leads to the following eigenvalue equation  
\begin{align}
     \hat{L}^2|n\rangle=\frac{1}{3}\big( 2|h_{1}(n)|^2 + 3\sqrt{3}\theta + \sqrt{3}Bg(n) + g^2(n)\big)|n\rangle. 
    \label{lensq-op2}
\end{align}  
Using Eq.(\ref{h1(n)}) and Eq.(\ref{g(n)-2}), we can reexpress the form of the eigenvalue of the length square operator as
\begin{align}
\hat{L}^2|n\rangle=\frac{1}{3}\big([2(n-\overline{n})+1]3\sqrt3\theta + \sqrt{3}Bg(\overline{n})+ g^{2}(\overline{n})
\big)|n\rangle. 
\label{lensq-op3}
\end{align}

From the expression for $g(\overline{n})$ deduced in Eq.(\ref{g(n-bar)}), it is evident that $g(\overline{n})$ is dependent on both $\tilde{n}$ and $\overline{n}$. In consequence, we also find that both $|h_1(n)|^2$ and $|h_2(n)|^2$ now depend on $\overline{n}$ and $\tilde{n}$ in addition to $n$. But, $\tilde {n}$ and $\overline{n}$ do not necessarily take fixed values and could assume different values such that Eq.(\ref{g-in}) is obeyed. In light of this, we now see that the state of the system should depend on $n$, $\tilde {n}$ and $\overline{n}$. As a result, we can infer that we will require three indices to express the system's state, although the eigenvalues of $\hat{L}^2$ depend only on $n-\overline{n}$ and $\tilde{n}-\overline{n}$. The state of the system can then better be represented as $|\tilde{n},\overline{n},n\rangle$. Now, employing Eq.(\ref{g(n-bar)}) in Eq.(\ref{lensq-op3}) and further simplifying, the eigenvalue of the length square operator thus emerges to be
\begin{align}
\hat{L}^2|\tilde{n},\overline{n},n\rangle = \big([2n\!-\!\tilde{n}\!-\!\overline{n}]\sqrt3\theta 
 \!+\! \tfrac{B^2}{4}(\tilde{n}-\overline{n})(\tilde{n}-\overline{n}+2)+ \tfrac{3{\theta}^2}{B^2}\big)|\tilde{n},\overline{n},n\rangle .
 \label{len-sq-final}
\end{align}
Fundamentally, we have derived the eigenvalues of
$\hat{L}^{2}$ by expressing them in relation to the eigenvalues associated with $\hat{a}_{+}\hat{a}_{-}$ and $\hat{b}$. The length is quantized in all directions in contrast to \cite{Balasundaram:2022esk}, where the length was quantized only in a plane. 

It can be seen that if $\theta$ is assigned the value of 0 in Eq.(\ref{len-sq-final}), the length-square equation reduces to the form of the angular momentum problem with the angular momentum quantum number $l$ analogous to the value $\frac{\tilde{n}-\overline{n}}{2}$ and $\hbar$ analogous to $B$.

Note that $n$ appears only in the first term which is square-bracketed in Eq.(\ref{len-sq-final}) and since $n$ takes the values from $\overline{n}$ to $\tilde{n}$, this first term in the cases when $n=\overline{n}$ and $n=\tilde{n}$ respectively involves $-[\tilde{n}-\overline{n}]$ and $[\tilde{n}-\overline{n}]$. Therefore, the minimum of $\hat{L}^2$ occurs when $n=\overline{n}$ since $\tilde{n}\geq \overline{n}$, and this minimum is given by the equation
\begin{align}
\hat{L}^2|\tilde{n},\overline{n},\overline{n}\rangle = \big(-[\tilde{n}\!-\!\overline{n}]\sqrt3\theta 
 \!+\! \tfrac{B^2}{4}(\tilde{n}-\overline{n})(\tilde{n}-\overline{n}+2)+ \tfrac{3{\theta}^2}{B^2}\big)|\tilde{n},\overline{n},\overline{n}\rangle .
 \label{len-sq-final2}
\end{align}
If $\tilde{n}=\overline{n}$, the eigenvalue of $\hat{L}^2$ is $\tfrac{3{\theta}^2}{B^2}$. But the eigenvalue of $\hat{L}^2$ can be lower than $\tfrac{3{\theta}^2}{B^2}$ if $\big(-[\tilde{n}\!-\!\overline{n}]\sqrt3\theta 
 \!+\! \tfrac{B^2}{4}(\tilde{n}-\overline{n})(\tilde{n}-\overline{n}+2)\big)<0$ i.e., if 
 \begin{equation}
     \tilde{n}<\overline{n}+(\tfrac{4\sqrt{3}\theta}{B^2}-2).
     \label{ntilde-nbar}
 \end{equation}
It is clear from Eq.(\ref{ntilde-nbar}) that if  $(\frac{4\sqrt{3}\theta}{B^2}-2)\leq 0$, then the inequality is violated and there cannot be a minimum lower than $\tfrac{3{\theta}^2}{B^2}$, i.e. if $\frac{\theta}{B^2} \leq \frac{1}{2\sqrt{3}}$, then the minimum eigenvalue of $\hat{L}^2$ is $\tfrac{3{\theta}^2}{B^2}$.
  
On the other hand, when $\tilde{n} = \overline{n}+m$, where $m>0$, the condition that the eigenvalue of $\hat{L}^2$ should be greater than or equal to zero leads to a real inequality relationship between $\tilde{n}$ and $\overline{n}$ only if $\frac{\theta}{B^2}\leq \frac{1}{4\sqrt{3}}$ which makes Eq.(\ref{ntilde-nbar}) inconsistent with $\tilde{n} = \overline{n}+m$ for positive $m$ and so $(-[\tilde{n}\!-\!\overline{n}]\sqrt3\theta 
 \!+\! \tfrac{B^2}{4}(\tilde{n}-\overline{n})(\tilde{n}-\overline{n}+2))$ cannot be less than zero in a consistent way. Therefore, the minimum eigenvalue of $\hat{L}^2$ is $\tfrac{3{\theta}^2}{B^2}$. This is also clear from Eq.(\ref{lensq-op3}) since the minimum of Eq.(\ref{g(n-bar)}) is $-\frac{3\theta}{B}$ and the minimum of $n-\overline{n}$ is 0.
\section{Degeneracy of States}

The dependency of the state of the system on more than one index can be better understood by the construction of another operator $\hat{K}$ that commutes with $\hat{a}_\pm$ and $\hat{b}$. Since $[\hat{L}^2,\hat{a}_\pm]= \pm2\sqrt{3}\theta\hat{a}_\pm$ and $[\hat{b},\hat{a}_\pm]=\mp\sqrt3\,B\,\hat{a}_\pm$, the linear combination of $\hat{L}^2$ and $\hat{b}$, $B \hat{L}^2 + 2\theta \hat{b}$, commutes with $\hat{a}_\pm$. But with little changes, we construct the following linear combination 
\begin{equation}
    \hat{K}= \hat{L}^2 + \frac{2\theta}{B}\hat{b} + \sqrt{3}\theta + \frac{3\theta^2}{B^2}
    \label{k-1}
\end{equation}
which also commutes with $\hat{a}_\pm$. 
Since $\hat{b}$ exhibits commutativity with $\hat{L}^2$, we can summarise the commutator relations as
\begin{align}
    [\hat{K}, \hat{a}_\pm] &= 0, \\
     [\hat{K}, \hat{b}] &= 0.
\end{align}
These equations, along with Eq.(\ref{ab-commutator}), form a set of equations analogous to the operator algebra involving $\hat{\mathcal{L}}^2$, $\hat{\mathcal{L}}_\pm$ and $\hat{\mathcal{L}}_z$ in the angular momentum problem in quantum mechanics. The eigenvalue of the operator $\hat{K}$ acting on the state $|\tilde{n},\overline{n},n\rangle$ can be evaluated from Eq.(\ref{len-sq-final}) and Eq.(\ref{k-1}) as
\begin{equation}
    \hat{K}|\tilde{n},\overline{n},n\rangle= \left(\frac{B^2}{4}(\tilde{n}-\overline{n})(\tilde{n}-\overline{n}+2) + \sqrt{3}\,\theta \right)|\tilde{n},\overline{n},n\rangle.
    \label{k-final}
\end{equation}

It is evident that the eigenvalue of the operator $\hat{K}$ is independent of $n$. Also, its eigenvalues do not change if $\tilde{n}$ and $\overline{n}$ are changed, keeping $\tilde{n}-\overline{n}$ fixed, resulting in huge degeneracy. 
\begin{figure}
\centering{\includegraphics[scale=0.45,clip, trim=1cm 11cm 1cm 2.5cm, width=0.5\textwidth]{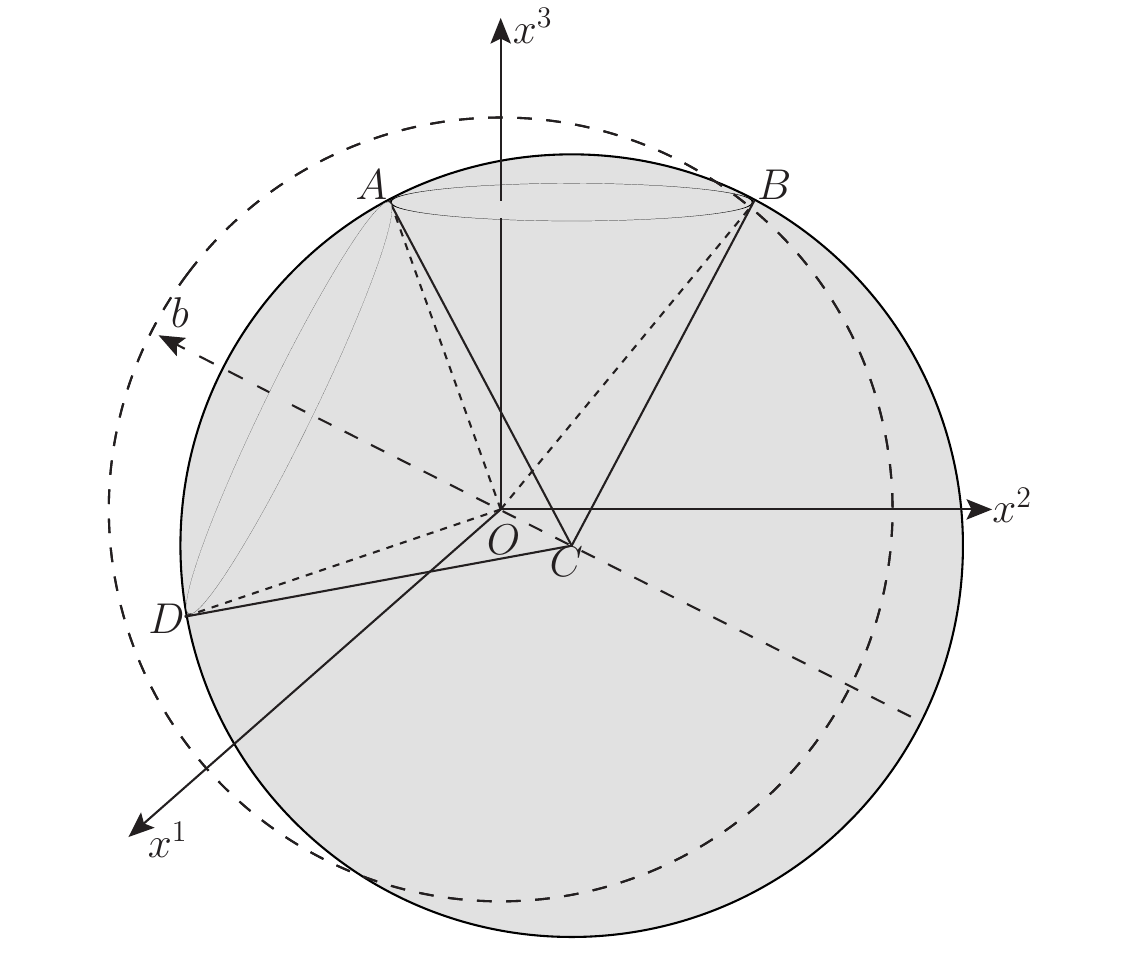}}
\caption{Degenerate states of operator $\hat{K}$ lying on the shaded-sphere} \label{sphere}
\end{figure}
\FloatBarrier
The physical meaning of $\hat{K}$ can be described in the following way. The Eq.(\ref{len-sq-AdaggerA}) represents the invariance of the length operator in going from the basis $\hat{X}$ to the basis $\hat{A}$. Any constant eigenvalue of $\hat{L}^2$ in Eq.(\ref{length-square1}) in the 3-D case would correspond to a sphere centered at the origin. In the basis $\hat{A}$, the constant eigenvalue of  $\hat{L}^2$ in Eq.(\ref{lensq-op1}) would represent the same sphere centered at the origin. Writing $\hat{K}$ in the form of the Eq.(\ref{lensq-op1}) leads to the expression
\begin{align}
    \hat{K} = \frac{1}{3}[2\hat{a}_+\hat{a}_- +3\sqrt{3} \theta+ \sqrt{3}B\hat{b}^\prime + {\hat{b}}^{\prime \, 2}], \label{K-as-shfted-len-sq}
\end{align}
where $\hat{b}^\prime = \hat{b}+\tfrac{3\theta}{B}$. If the constant eigenvalue of Eq.(\ref{lensq-op1}) represents a sphere centered at the origin, then the constant eigenvalue of Eq.(\ref{K-as-shfted-len-sq}) represents a sphere shifted along $\hat{b}$ axis by the amount $-3\theta/B$. In Figure \ref{sphere}, the sphere corresponding to a constant eigenvalue of $\hat{L}^2$ is represented by the dashed circle centered at $O$ and the shifted-sphere is represented as the shaded sphere centered at $C$. The degenerate eigenstates of $\hat{K}$ lie on the surface of this shifted sphere, but these states will have different eigenvalues for $\hat{L}^2$ since $\hat{L}^2$ is measured from the origin $O$. While $CA$ and $CB$ corresponding to the eigenvalues of $\hat{K}$ are equal, $OA$ and $OB$ corresponding to the eigenvalues of $\hat{L}^2$ are not equal. For example, the states $|\tilde{n},\overline{n},n\rangle=|8,2,n\rangle$ with different $n$'s have different eigenvalues for $\hat{L}^2$ in Eq.(\ref{len-sq-final}), but they have the same eigenvalue for $\hat{K}$ in Eq.(\ref{k-final}). But the states $|8,2,5\rangle$ and $|9,3,6\rangle$ have the same eigenvalue for $\hat{L}^2$ and the same eigenvalue for $\hat{K}$. Such a common set of degenerate states of $\hat{L}^2$ and $\hat{K}$ lie on a circle perpendicular to the $\hat{b}$ axis since the shift happens along the $\hat{b}$ axis. 

In the basis $(\hat{x}^1,\hat{x}^2,\hat{x}^3)$, the shifted sphere corresponds to the shift along all three directions. In other words, if $d^i$ denotes the shift along the $x^i$ direction, then defining the operator $\hat{y}^i= \hat{x}^i+d^i$ such that $B^{ij}_{\phantom{ij}k}d^k=\theta^{ij}$ in Eq.(\ref{commutator2}) would lead to the Lie structure $[\hat{y}^i,\hat{y}^j]=i B^{ij}_{\phantom{ij}k}\hat{y}^k$. Although this structure would lead to the quantization of $(\hat{y}^1)^2+(\hat{y}^2)^2+(\hat{y}^3)^2$, our length operator is different from this and in terms of $\hat{y}^i$, it is  $(\hat{y}^1-d^1)^2+(\hat{y}^2-d^2)^2+(\hat{y}^3-d^3)^2$. While the quantization method using $\hat{y}^i$ would employ raising and lowering operations along $\hat{y}^3$ direction and keep the eigenvalues of $(\hat{y}^1)^2+(\hat{y}^2)^2+(\hat{y}^3)^2$ fixed, it can be easily shown that $\hat{y}_\pm=\hat{y}^1\pm i\hat{y}^2$ neither raises/lowers the eigenstates of $\hat{L}^2$ nor it commutes with $\hat{L}^2$, calling for the approach that starts with the Eq.(\ref{length-lad-comm}) to look for the properly oriented $\hat{a}_\pm$ in place of $\hat{y}_\pm$.

\section{Conclusion}
In conclusion, we have explored length quantization in the context of noncommutative spaces with position-dependent noncommutativity. Building upon the formalism similar to the quantum harmonic oscillator and angular momentum problems, we have constructed ladder operators and derived the operator corresponding to the length-square in terms of these ladder operators. This investigation has been specifically applied to the case of a 3-dimensional space where the noncommutativity parameter involved a combination of canonical/Weyl-Moyal type and Lie-type. 

We have found that the length quantization in this scenario leads to distinct, discrete eigenvalues for the length-square operator. The ladder operators drive the behavior of eigenvalues, resulting in the quantization of length not only within a plane but also along a direction normal to that plane. The derived ladder operators and their commutation relations have enabled us to construct a comprehensive operator for the square of length. This operator yields a structured ladder of eigenstates that are simultaneously eigenstates of both the length-square operator and certain combinations of ladder operators. Through this approach, we have given the formalism of how quantization occurs within spaces with position-dependent noncommutativity. Furthermore, we have identified the ground state and explored the maximum and minimum possible values of the quantum numbers associated with the ladder operators, thereby defining the range of valid eigenstates.

The study of degenerate states constitutes a significant facet of our research inquiry. In our analysis, we have examined the implications of position-dependent noncommutativity on the degeneracy of states within the framework of the length quantization problem. By introducing an operator $K$, we have unveiled the distinct conditions under which degenerate states emerge, shedding light on the intricate balance between spatial geometry and quantum behavior. 

The quantization of length in noncommutative spaces with position-dependent noncommutativity opens up new avenues for exploring the fundamental nature of spacetime and its implications for physical theories. Our findings may inspire further investigations into the intriguing interplay between geometry and quantum algebra in noncommutative spacetime settings. We have presented only the quantization of a bare fundamental geometric element, i.e., length. The construction of a field theory or mechanics with the underlying quantized geometric element is beyond the scope of this work. 

\end{document}